\documentclass[journal,a4paper]{IEEEtran}

\usepackage{graphicx}
\usepackage{color}
\usepackage[normalem]{ulem}

\usepackage{wasysym} 

\begin{document}
%
\title{Reactive Magnetron Sputter Deposition of Superconducting Niobium Titanium Nitride Thin Films with Different Target Sizes}
%
%
%

\author{B.~G.~C.~Bos,
        D.~J.~Thoen,
        E.~A.~F.~Haalebos,
        P.~M.~L.~Gimbel,
        T.~M.~Klapwijk,
        J.~J.~A.~Baselmans,
        and~A.~Endo
\thanks{B.G.C. Bos, D.J. Thoen, P.M.L. Gimbel, T.M. Klapwijk and A. Endo are with the Kavli Institute of NanoScience, Faculty of Applied Sciences, Delft University of Technology, Lorentzweg 1, 2628 CJ Delft, The Netherlands. e-mail: (see http://qn-med.tudelft.nl/).}
\thanks{D.J. Thoen, J.J.A. Baselmans and A. Endo are with the Department of Electrical Engineering, Faculty of Electrical Engineering, Mathematics and Computer Science (EEMCS), Delft University of Technology, Mekelweg 4, 2628 CD Delft, The Netherlands. e-mail: (see http://terahertz.tudelft.nl/).}
\thanks{E.A.F. Haalebos and J.J.A. Baselmans are with the Netherlands Institute for Space Research (SRON), Sorbonnelaan 2, 3584 CA Utrecht, The Netherlands.}
\thanks{T.M. Klapwijk is with the Physics Department, Moscow State Pedagogical University, 119991 Moscow, Russia.}
}

%
%

%

\maketitle

\begin{abstract}
The superconducting critical temperature ($T_\mathrm{c} >$ 15 K) of niobium titanium nitride (NbTiN) thin films allows for low-loss circuits up to 1.1 THz, enabling on-chip spectroscopy and multi-pixel imaging with advanced detectors. The drive for large scale detector microchips is demanding NbTiN films with uniform properties over an increasingly larger area. This article provides an experimental comparison between two reactive d.c. sputter systems with different target sizes: a small target ($\diameter$100 mm) and a large target (127 mm $\times$ 444.5 mm). This article focuses on maximizing the $T_\mathrm{c}$ of the films and the accompanying $I$-$V$ characteristics of the sputter plasma, and we find that both systems are capable of depositing films with $T_\mathrm{c} >$ 15 K. The resulting film uniformity is presented in a second manuscript in this volume. We find that these films are deposited within the transition from metallic to compound sputtering, at the point where target nitridation most strongly depends on nitrogen flow. Key in the deposition optimization is to increase the system's pumping speed and gas flows to counteract the hysteretic effects induced by the target size. Using the $I$-$V$ characteristics as a guide proves to be an effective way to optimize a reactive sputter system, for it can show whether the optimal deposition regime is hysteresis-free and accessible.
\end{abstract}


\section{Introduction}
Reactive sputter deposition is a physical vapour deposition technique, where a metallic target is sputtered by ions of an inert element from a plasma while a flow of reactive gas passes the the whole system, including the target, plasma and substrate \cite{ohring}\cite{deplamahieu}. For the fabrication of niobium titanium nitride (NbTiN), an alloy target of niobium and titanium is used in combination with argon and nitrogen as inert and reactive gasses respectively.

Current fabrication of NbTiN thin films at the Kavli Institute of NanoScience in Delft, the Netherlands, is performed using a Nordiko 2000 reactive sputter deposition machine. Yet the new astronomical sub-millimeter detectors A-MKID \cite{baryshev} and DESHIMA \cite{endo} require a uniform thin film over the surface of a complete \diameter100 mm wafer. Uniformity on so large a surface cannot be obtained with the small \diameter100 mm target of the Nordiko 2000, fit with an even smaller \diameter60 mm magnet. As the size of the target and specifically the size of the erosion track is expected to influence the thickness profile of the film, we make a switch to the Evatec LLS801 at SRON Utrecht. This system houses a large 127 mm $\times$ 444.5 mm target on which the erosion track covers almost 70\% of the surface. Although a small target is more easily applicable, for the required gas flows and power are lower, this larger target is expected to result in better deposition uniformity on a large substrate concerning thickness, critical temperature ($T_\mathrm{c}$) and other material quality factors. The aim of this article is to provide an empirical insight in the shift from a small-target reactive sputter machine to a large-target reactive sputter machine, while maintaining a high $T_\mathrm{c}$ of 15 K in the center of the deposited films. This insight is granted by exploring the $I$-$V$ characteristics of both sputter machines.

\section{Theory}

A deposited NbTiN thin film with good superconducting properties will only develop under the right stoichiometry \cite{benvenuti} and microstructural growth \cite{thornton1986}\cite{iossad}. To get an idea of these processes, it is crucial to understand the plasma characteristics and the interaction between the plasma and the target. Empirically, the plasma can be probed by the $I$-$V$ relationship of the system. According to Depla \textit{et al.} \cite{depla2006}, the $I$-$V$ characteristics of a reactive sputter system can be formulated as

\begin{equation}
\label{iv}
I = kV^{n},
\end{equation}

where $k$ and $n$ are constants that depend on parameters of the experimental set-up, such as gas-target combination, target condition, magnetic field and system geometries. Thornton \cite{thornton1978} states that the minimum target voltage to sustain a plasma is given by

\begin{equation}
V_{min}=W_{0}/{\Gamma_{i}\epsilon_{i}\epsilon_{e}}.
\end{equation}

Here $W_{0}$ is the argon ionization energy (around 16 eV for Ar$^{+}$ \cite{CRC}), $\Gamma_{i}$ represents the effective number of secondary electrons released from the target (cathode) per primary free electron, $\epsilon_{i}$ is the fraction of ions that collide with the target and $\epsilon_{e}$ denotes the fraction of the secondary electron's energy $eV_{target}$ that has been transferred into exciting argon atoms before the electron is captured by the anode. For magnetron sputter deposition, $\epsilon_{i}$ and $\epsilon_{e}$ do not vary much from unity. Buyle \textit{et al.} \cite{buyle} further distinguished $\Gamma_{i}=\gamma_{ISEE}\times E$, where $\gamma_{ISEE}$ is the ion-induced secondary electron emission coefficient and $E$ is the effective gas ionization probability.

The parameters that define the target voltage are mostly determined by the target-plasma characteristics. $W_{0}$ depends on the chosen inert working gas, $\gamma_{ISEE}$ is a property of the target metal and the fraction of its surface covered with compound \cite{depla2007} and $E$ is especially strongly dependent on the gas pressure. Therefore, the $I$-$V$ relationship is suitable to provide a good empirical insight in the behaviour of the plasma and its interactions with the target, which eventually form the basis of the thin film growth and thus primarily define the properties of the deposited thin film.

\section{Experimental Set-up}
The system with the small target is the commercial Nordiko 2000 from Nordiko Technical Services Ltd. It houses a round Nb-Ti target of \diameter 100 mm and 6 mm thickness. The LLS801 is originally an industry-based Balzers LLS801, which is refurbished for research purposes by Evatec. It contains a large rectangular Nb-Ti target with dimensions 127 mm $\times$ 444.5 mm $\times$ 12 mm (w$\times$l$\times$th) of type AK 517, equipped with a magnetron type B. In contrast to the Nordiko, the substrate is mounted on a rotatable drum with a diameter of 670 mm. This allows the drum to oscillate in front of the target in the direction of the target's short side to improve the uniformity even further. Both targets are made from the same batch, consisting of 69 at.\% niobium and 31 at.\% titanium (table \ref{specs}), which suffices for a high $T_\mathrm{c}$ \cite{dileo}, and are purchased at Thermacore Inc. while fabrication took place at their Materials Technology Division in Pittsburgh, PA, USA. Particle analysis shows that magnetic impurities have been reduced to 30 parts per million (PPM) for oxygen, and less than the ICP-AES measurement limit for iron $<$23.1 PPM, chromium $<$9.23 PPM and nickel $<$4.62 PPM. The base pressures of the sputter chambers are limited by water vapour at total pressures of $2\cdot10^{-5}$ Pa for the Nordiko 2000 and $3\cdot10^{-5}$ Pa for the LLS801, which suffices to reduce contaminants in the film \cite{nakano}. The argon gas purity is 99.9999\% for the Nordiko 2000 and 99.999\% for the LLS801, where both systems have nitrogen gas purities of 99.9999\%. In both sputter machines the substrate is grounded and the power supplies are operated in d.c. mode and power biased. The substrate holder of the Nordiko 2000 is actively cooled, where the substrate holder of the LLS801 is not. Further specifications of both systems are visible in table \ref{specs}.

\begin{table}
\renewcommand{\arraystretch}{1.3}
\caption{Specifications of the Nordiko 2000 and LLS801}
\label{specs}
\centering
\begin{tabular}{c c c}
\hline
\bfseries Parameter & \bfseries Nordiko 2000 & \bfseries LLS801\\
\hline
Target geometry & \diameter100 mm & 127 mm $\times$ 444.5 mm \\
Target erosion track area & 30 cm$^{2}$ & 330 cm$^{2}$ \\
Target Nb:Ti at. ratio & 69:31 & 69:31 \\
Target-substrate distance & 80 mm & 90 mm \\
Maximum power & 550 W & 5.0 kW \\
Maximum Ar flow & 100 sccm & 500 sccm \\
Maximum N$_{2}$ flow & 20 sccm & 100 sccm \\
Maximum pump speed (air) & 1500 L/s & 2200 L/s  \\
\hline
\end{tabular}
\end{table}

In order to obtain the $I$-$V$ relationship of a system, a preset argon input flow is chosen. For differently set nitrogen input flows, the applied power will be swept from low to high power and backwards. After each power change, the system has a one minute delay time for the plasma to find a stable current and voltage. The setting of the throttle valve in front of the cryogenic pump will determine the gas pressure in the chamber.

A cross section of the $I$-$V$ curves measured with this method, can be obtained by conducting a nitrogen sweep. In this case the applied power and argon input flow are kept at a constant level and the nitrogen input flow is varied from minimum to maximum. After every step in nitrogen input flow, a one minute delay is used.

 The focus of this article is on the $T_\mathrm{c}$ at the center position of the superconducting thin film and on the place these depositions take in the $I$-$V$ traces of the systems. The uniformity obtained in the LLS801 will not be discussed further in this article, but is in focus elsewhere \cite{thoen}. Films are sputtered to be of a thickness of 500 nm. The $T_\mathrm{c}$ is determined by cooling the thin films in a liquid helium dewar, while continuously measuring a four probe resistance. Due to the usage of a liquid helium bath, the lower limit of the temperature in our measurement is 4.2 K. For films that have $T_\mathrm{c}<$ 4.2 K, we can only set an upper limit. The temperature can be regulated with a small resistive heat source. The superconducting transition tends to show some hysteresis in temperature in the order of 1\% of the $T_\mathrm{c}$, as the temperature of the film slightly lags the measured temperature. The $T_\mathrm{c}$'s mentioned in this work are the average of the upward and downward transition temperature. Furthermore, the $T_\mathrm{c}$ is defined as the temperature where the resistance of the thin film is half the resistance at 20 K.

\section{Results and discussion}

\begin{figure}
\centering
\includegraphics[width=0.48\textwidth]{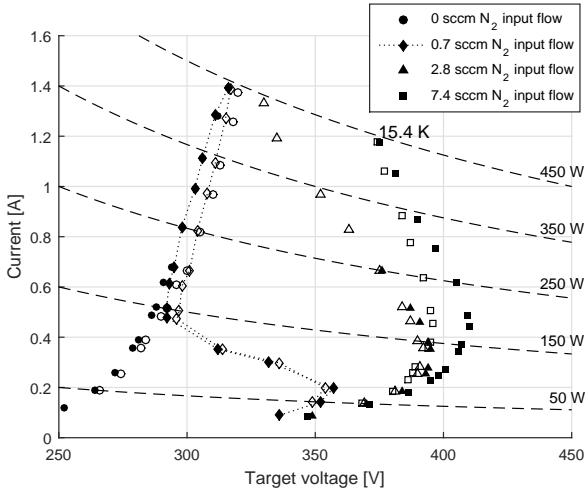}
\caption{Nordiko 2000 $I$-$V$ relationships. The argon input flow is 100.3 sccm and the argon partial pressure 0.7 Pa. The point of a high $T_\mathrm{c}$ deposition is indicated. Isopower curves are depicted by the dashed lines. Upward power sweeps are indicated by the filled symbols, downward power sweeps by the unfilled symbols.}
\label{nordiko}
\end{figure}

The Nordiko 2000 is a well explored machine that can fabricate NbTiN thin films with a $T_\mathrm{c}$ up to 15.7 K when using a sputter pressure of 0.5 Pa, a room temperature resistivity as low as 87 $\mu\Omega$ cm \cite{thoen} and microwave resonator quality factors over $10^6$ for a 300 to 550 nm thickness \cite{barends}. In Fig. \ref{nordiko} the $I$-$V$ characteristics are shown for an argon partial pressure of 0.7 Pa. Here, the input argon flow is 100.3 sccm and curves for different input nitrogen flows are depicted. Furthermore, the position in the $I$-$V$ spectrum where films with high $T_\mathrm{c}$ are obtained, is included in the figure. 

The zero nitrogen curve in Fig. \ref{nordiko} follows the relation given by equation \ref{iv}. This curve is the metallic branch, where the target is fully metallic. For low power the non-zero nitrogen curves follow the same relationship, but for different constants $k$ and $n$. This is the compound branch, because the target is fully covered with compound NbTiN here. At a certain applied power, the curve will smoothly deviate from the compound branch towards the metallic branch. The required power for this transition is higher for a larger input nitrogen flow, because more nitrogen has to be sputtered from the target. The nitridation of the target increases the target voltage, due to the decrease of the ion-induced secondary electron emission coefficient $\gamma_{ISEE}$ \cite{mientus}. Partial nitridation of the target and the accompanied raise in target voltage have a positive effect on the $T_\mathrm{c}$ \cite{matsunaga}. The highest sputter rates of films deposited at 440 W in a nitrogen rich environment are about 2.8 $\mathrm{nm\ s^{-1}}$;  high $T_\mathrm{c}$ films are sputtered at about 1.35 $\mathrm{nm\ s^{-1}}$. 

\begin{figure}
\centering
\includegraphics[width=0.48\textwidth]{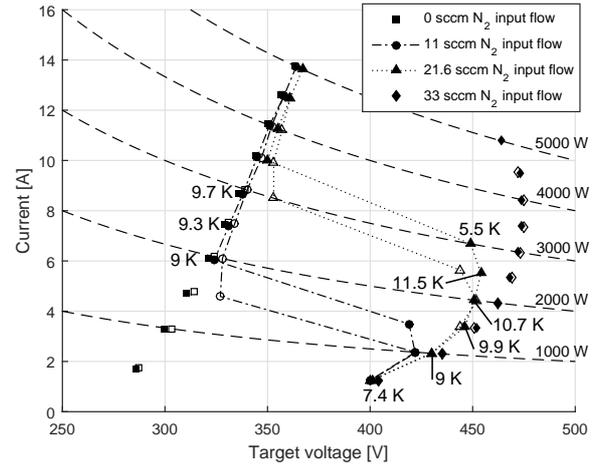}
\caption{LLS801 $I$-$V$ relationships in the low flow regime. The argon input flow is 55.4 sccm and the argon partial pressure 0.6 Pa. Deposited thin films have an added $T_\mathrm{c}$ value. Isopower curves are depicted by the dashed lines. Upward power sweeps are indicated by the filled symbols, downward power sweeps by the unfilled symbols.}
\label{llslowflow}
\end{figure}

For the first round of experiments in the LLS801, its parameters are adjusted to operate around the same values as the Nordiko 2000. In Fig. \ref{llslowflow} the $I$-$V$ curves are presented for an input argon flow of 55.4 sccm, different input nitrogen flows and an argon partial pressure of 0.6 Pa. Again, for some $I$-$V$ values thin films are fabricated and $T_\mathrm{c}$ is obtained.

Some differences are apparent between Fig. \ref{nordiko} and Fig. \ref{llslowflow}. Firstly, the deviation of the curves with constant nitrogen input flow from the compound branch towards the metallic branch ends with a sudden leap, which indicates hysteretic behaviour of the system. According to the Berg-model for reactive sputter deposition \cite{berg2005}, this hysteresis can be explained by the increase in target size, relative to the Nordiko 2000, which intensifies the effect of nitrogen consumption of the target on the nitrogen partial pressure. Secondly, the $T_\mathrm{c}$'s obtained from sputtered films from the system do not exceed 12 K. Higher T$_{c}$ is obtained in the Nordiko 2000 at a point within the transition from the metallic to the compound branch, a point that is not accessible in the LLS801 due to hysteresis.

\begin{figure}
\centering
\includegraphics[width=0.48\textwidth]{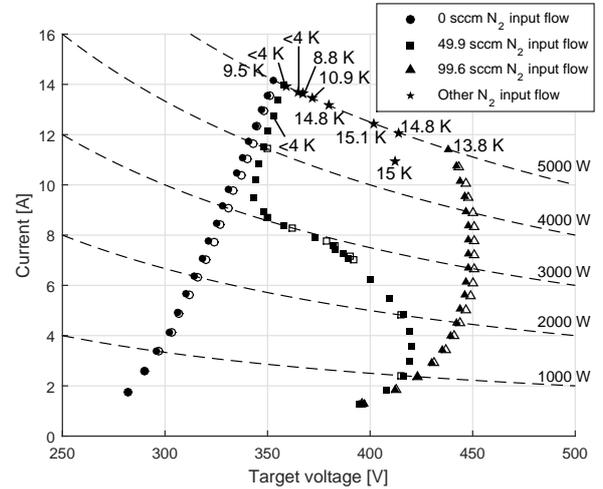}
\caption{LLS801 $I$-$V$ relationships in the high flow regime. The argon input flow is 400.8 sccm and the argon partial pressure 0.7 Pa. Deposited thin films have an added $T_\mathrm{c}$ value. Isopower curves are depicted by the dashed lines. Upward power sweeps are indicated by the filled symbols, downward power sweeps by the unfilled symbols.}
\label{llshighflow}
\end{figure}

The Berg-model also presents a solution to eliminate the hysteresis by increasing the system's pumping speed. Therefore, for the second round of experiments, the throttle valve of the LLS801 is completely opened in order to reach the maximum pumping speed of the cryogenic pump. The argon input flow is raised to 400.8 sccm to obtain an argon partial pressure of 0.7 Pa, while the nitrogen input flow is elevated to match the nitrogen consumption on the target. The resulting $I$-$V$ curves of the power sweeps at constant nitrogen flow are depicted in Fig. \ref{llshighflow}, together with the $T_\mathrm{c}$ values of some sputtered thin films.

When sputtering in the high flow regime of the LLS801 (fig. \ref{llshighflow}), the smooth $I$-$V$ curves visible in Fig. \ref{nordiko} return. Every $I$-$V$ position becomes stably accessible and good NbTiN thin films can be fabricated with $T_\mathrm{c} >$ 15 K in the transition from the metallic to the compound branch. The films with a $T_\mathrm{c}$ of 14.8 K or higher are sputtered with a nitrogen input flow between 80 and 86 sccm. 

\begin{figure}
\centering
\includegraphics[width=0.48\textwidth]{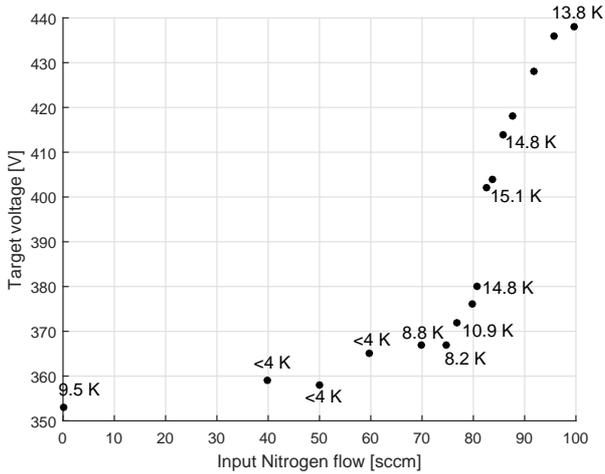}
\caption{LLS801 nitrogen sweep. The argon input flow is 400.8 sccm, the argon partial pressure 0.7 Pa and the applied power 5 kW. Deposited thin films have an added $T_\mathrm{c}$ value.}
\label{llsn2sweep}
\end{figure}

A better understanding on where exactly between the metallic branch and the compound branch the film with highest $T_\mathrm{c}$ can be sputtered, is given by imaging the 5 kW isopower line visible in Fig. \ref{llshighflow}. In order to do so, the applied power is set to 5 kW and the nitrogen input flow is varied. These measurements result in the V-N$_{2}$ curve shown in Fig. \ref{llsn2sweep}, where the obtained $T_\mathrm{c}$'s from a few films are also included. The maximum deposition rate in a nitrogen rich plasma at 5.0 kW is about 5.4 $\mathrm{nm\ s^{-1}}$, films with the highest $T_\mathrm{c}$ are sputtered at a rate of 4.8 $\mathrm{nm\ s^{-1}}$.

It is visible in Fig. \ref{llsn2sweep} that the optimal NbTiN thin film is sputtered for a certain nitrogen input flow, where the target voltage is most strongly dependent on the nitrogen input flow or, in other words, where the double derivative equals zero. This is the point where the target most strongly binds nitrogen, for a saturated Nb-Ti target requires a higher target voltage to sustain a plasma. At exactly this point the Berg-model \cite{berg2005} also predicted the optimal films to be sputtered, because here the deposited film is fully stoichiometric, whereas the target is not. This has been experimentally verified \cite{bedorf}\cite{sproul}.

As mentioned earlier, the reason why this high $T_\mathrm{c}$ regime is not accessible in the case of the LLS801 with low flows (fig. \ref{llslowflow}), is hysteresis. In an hysteretic reactive sputter system, at a certain input flow of reactive gas, both the target and substrate are either under- or oversaturated with reactant \cite{guttler}. This will prevent any access to the "sweet spot" of reactive sputtering, where the thin film is completely saturated with reactant and the target is not.

\section{Conclusion}
A reactive sputter deposition machine is a complex physical system, where electrical, geometrical and material parameters together define the plasma behavior, the plasma-target interaction and finally the deposition onto the substrate. We show that by using the shape of $I$-$V$ curves as a guide, we can easily examine the behavior of the system to adapt film properties to our need. The $I$-$V$ curves are a tool to compare various set-ups with different target size, for it provides quantitative feedback depending on every internal parameter and setting.

Furthermore, NbTiN thin films with high $T_\mathrm{c}$ can only be sputtered in the transition of a metallic target towards a saturated target. This transition should therefore be accessible and free of hysteresis. The system parameters required for such behaviour are highly dependent on the target size, which primarily defines the nitrogen consumption. Eliminating hysteresis by using an appropriate pumping speed and gas flows, opens up the possibility to fabricate stoichiometric NbTiN with $T_\mathrm{c} >$ 15 K.

\section*{Acknowledgment}

The authors would like to thank Marcel Bruijn and Vignesh Murugesan, members of the SRON clean room staff, for their support in the fabrication and processing at SRON. This research was supported by the NWO (Netherlands Organisation for Scientific Research) through the Medium Investment grant (614.061.611). AE was supported by the NWO Vidi grant (639.042.423). TMK was supported by the Ministry of Science and Education of Russia under Contract No. 14.B25.31.0007 and by the European Research Council Advanced Grant No. 339306 (METIQUM).

\ifCLASSOPTIONcaptionsoff
  \newpage
\fi



\bibliographystyle{IEEEtran}

\bibliography{ivPaper_v01}

\end{document}